\documentclass[12pt,preprint,prd,showpacs,showkeys]{revtex4}
\usepackage{amsmath,amssymb}
\usepackage{graphicx}

\begin{document}
\title{Production of Neutrinos and Secondary Electrons in Cosmic Sources}
\author{C.-Y. Huang\footnote{Corresponding author: Tel.: +1-515-2945062; fax: +1-515-2946027\\
                        {\it E-mail address:} huangc@iastate.edu (C.-Y. Huang)}}
\author{M. Pohl}
\affiliation{Department of Physics and Astronomy, Iowa State University, Ames, IA 50011}

\def\AstroPartPhys#1#2#3{Astropart. Phys.~#1~(#2)~#3.}
\def\AstropartPhys#1#2#3{Astropart. Phys.~#1~(#2)~#3.}
\def\AA#1#2#3{Astron. Astrophys.~#1~(#2)~#3.}
\def\APSS#1#2#3{Astrophys. \& Space Science~#1~(#2)~#3.}
\def\APJ#1#2#3{Astrophys. J.~#1~(#2)~#3.}
\def\APJSuppl#1#2#3{Astrophys.J.Suppl.~#1~(#2)~#3.}
\def\MNRAS#1#2#3{Mon. Not. R. Astron. Soc.~#1~(#2)~#3.}
\def\MNRASLett#1#2#3{Mon. Not. R. Astron. Soc. Lett.~#1~(#2)~#3.}
\def\PRD#1#2#3{Phys. Rev. D~#1~(#2)~#3.}
\def\PRL#1#2#3{Phys. Rev. Lett.~#1~(#2)~#3.}
\def\PLB#1#2#3{Phys. Lett. B~#1~(#2)~#3.}
\def\NPB#1#2#3{Nucl. Phys. B~#1~(#2)~#3.}
\def\NPBProcSuppl#1#2#3{Nucl. Phys. B (Proc. Suppl.)~#1~(#2)~#3.}
\def\NPBProcSupplAB#1#2#3{Nucl. Phys. B (Proc. Suppl.)~#1~A,B~(#2)~#3.}
\def\JPhysG#1#2#3{J. Phys. G.~#1~(#2)~#3.}
\def\JPhysGNuclPartPhys#1#2#3{J. Phys. G: Nucl. Part. Phys~#1~(#2)~#3.}
\def\Nature#1#2#3{Nature~#1~(#2)~#3.}
\def\PR#1#2#3{Phys. Rev.~#1~(#2)~#3.}
\def\PASJ#1#2#3{PASJ~#1~(#2)~#3.}
\def\PhysRept#1#2#3{Phys. Rept~#1~(#2)~#3.}
\def\AnnuRevAstroAstrophys#1#2#3{Annu. Rev. Astron. Astrophys.~#1~(#2)~#3.}
\def\Nature#1#2#3{Nature~#1~(#2)~#3.}
\def\JCAP#1#2#3{JCAP~#1~(#2)~#3.}
\def\SpaceSciRev#1#2#3{Space Sci. Rev.~#1~(#2)~#3.}
\def\SovJNuclPhys#1#2#3{Sov. J. Nucl. Phys.~#1~(#2)~#3.}
\def\AIP#1#2#3{AIP Conf. Proc.~#1~(#2)~#3.}
\def\ICRCYr#1#2{Proc.~{#1}th Int. Cosmic Ray Conf. #2.}
\def\ICRCVol#1#2#3#4{Proc.~{#1}th Int. Cosmic Ray Conf. #2. (#3) #4.}
\def\NIMA#1#2#3{Nucl. Instrum. Methods Phys. Res. A~#1~(#2)~#3.}
\def\RepProgPhys#1#2#3{Rep. Prog. Phys.~#1~(#2)~#3.}
\def\mkp{\bf}
\def\huangc{\bf}

\begin{abstract}
We study the individual contribution to secondary lepton production in hadronic interactions 
of cosmic rays (CRs) including resonances and heavier secondaries. For this purpose we use
the same methodology discussed earlier \cite{Huang07}, namely the Monte Carlo particle collision code 
DPMJET3.04 to determine the multiplicity spectra of various secondary particles with 
leptons as the final decay states, that result
from inelastic collisions of cosmic-ray protons and Helium nuclei with
the interstellar medium of standard composition. By combining the simulation results with 
parametric models for secondary particle (with resonances included) for incident 
cosmic-ray energies below a few GeV, where DPMJET appears unreliable, we thus derive production matrices 
for all stable secondary particles in cosmic-ray interactions with energies up to about 10 PeV.

We apply the production matrices to calculate the radio synchrotron radiation of secondary 
electrons in a young shell-type SNR, RX J1713.7-3946, which is a measure of the age, the spectral
index of hadronic cosmic rays, and most importantly the magnetic field strength. We find that the
multi-mG fields recently invoked to explain the X-ray flux variations are unlikely to extend over a large 
fraction of the radio-emitting region, otherwise the spectrum of hadronic cosmic rays in the energy
window 0.1-100~GeV must be unusually hard.

We also use the production matrices to calculate the muon event rate in an IceCube-like detector
that are induced by muon neutrinos from high-energy $\gamma$-ray 
sources such as RX J1713.7-3946, Vela Jr. and MGRO J2019+37. At muon energies of a few TeV,
or in other word, about 10~TeV neutrino energy, an accumulation of 
data over about five to ten years would allow testing the hadronic origin of TeV $\gamma$-rays.
\end{abstract}
\maketitle
\section{Introduction}\label{Section:Introduction}
It is known that hadronic cosmic rays interacting with the interstellar medium (ISM) 
can produce $\gamma$-rays and leptons. 
The principal production mechanisms for $\gamma$-rays in high-energy astrophysics 
are inelastic CR+ISM interactions with subsequent decays of the secondaries 
(mostly the neutral $\pi^0$ and $\eta$) into $\gamma$-rays. The decays of 
charged $\pi^\pm$ mesons and other secondaries will result in the production of high-energy 
neutrinos and secondary leptons, electrons and positrons. In GeV-scale Galactic cosmic rays, 
the secondary electrons and positrons comprise around 20\% of the total electron flux and
significantly contribute to the non-thermal electromagnetic 
radiation of $\gamma$-ray sources.
The flux of cosmic-ray neutrinos is determined by the cosmic-ray 
abundance and the production cross sections 
of the parent particles ($\pi^\pm,~K^\pm,~K^0_S,~K^0_L,~\Sigma^\pm,~\Sigma^0,~\Lambda,~\bar{\Lambda}$) 
in the collisions of cosmic-rays with the interstellar gas nuclei. 

The purpose of this work is to carefully calculate the lepton production in cosmic-ray interactions. 
For this purpose we use the high-energy physics event 
generator DPMJET-III \cite{Roesler00} to simulate all secondary productions in both p-generated and 
He-generated interactions, similar to earlier work on hadronic $\gamma$-ray production
\cite{Huang07}. Our study includes direct lepton production 
and the decays of all relevant secondary particles with leptons as the final decay products. 
For the cosmic rays, protons and helium 
nuclei are taken into account. We assume the composition of the ISM as 90\% protons, 
10\% helium nuclei, 0.02\% carbon, and 0.04\% oxygen. At energies below 20 GeV, 
where DPMJET appears unreliable, we combine the simulation results with parametric models for 
$\pi^\pm$ and $\pi^0$ production \cite{Blattnig00,Kamae06} that include the production
of the resonances $\Delta$(1232) and $\Delta$(1600) and their subsequent decays, thus deriving a lepton 
production matrix for cosmic rays with energies up to about 10 PeV that can be easily 
used to interpret the spectra of cosmic lepton sources.  

We use our production matrices to calculate the CR-induced electron/positron production in young shell-type supernova remnants (SNR), for which the neutrino fluxes 
are also calculated. Observations of non-thermal X-ray synchrotron radiation from the SNRs SN~1006 
\cite{Koyama95}, RX J1713.7-3946 \cite{Koyama97}, IC 443 \cite{Keohane97,Slane99}, Cas A \cite{Allen97}, 
and RCW 86 \cite{Borkowski01} support the hypothesis 
that the Galactic cosmic-ray electrons may be accelerated predominantly in SNR,
although other acceleration sites of high-energy electrons might exist in the Galaxy. The existence
of electrons in SNRs with an energy up to about 100 TeV implies TeV-scale $\gamma$-ray emission
\cite{Pohl96} besides that possibly produced by hadronic cosmic rays, and TeV-scale gamma-rays 
have indeed been detected from the SNRs 
Vela Junior \cite{Aharonian05VelaJr} and RX J1713.7-3946 \cite{Aharonian06RXJ1713.7-3946}.
Although the hadronic interpretation is often favored to explain the $\gamma$-ray 
emission from SNRs such as RX J1713.7-3946
\cite{Aharonian06RXJ1713.7-3946}, there is still no direct observational evidence for nucleon 
acceleration in SNRs \cite{Reimer02, Huang07}. 
Currently, both hadronic and leptonic models can describe the observed TeV $\gamma$-rays in 
different astrophysical objects, but 
have problems of similar magnitude \cite{Katz07}. 

RX J1713.7-3946 is a unique SNR in the sense that its X-ray emission is strongly dominated by a 
non-thermal component, which is presumed to be synchrotron
radiation of ultrarelativistic electrons (see references \cite{Hiraga05, Drury01,Uchiyama07} and 
references therein for details). The recent broadband X-ray spectroscopy
performed with Chandra \cite{Uchiyama07} and the Suzaku experiment \cite{Takahashi07} provides 
evidence of very effective acceleration of particles in the shell of 
RX J1713.7-3946 and also reveals that the X-ray emission from two compact regions is variable in flux, 
which the authors interpret as indication of multi-mG magnetic fields. If this conclusion is correct and applies to the entire remnant, then 
the observed TeV-band $\gamma$-ray emission must be of hadronic origin. Here we show
that the synchrotron flux from the secondary electrons in RX~J1713.7-3946 would
in fact exceed the observed radio flux from this object, if multi-mG magnetic fields would permeate
the remnant, so a magnetic field of such magnitude will likely exist at most in a small fraction of emission
region.

In a hadron accelerator, TeV neutrinos should be produced in roughly the same number as TeV $\gamma$-rays. 
The Universe is generally more transparent for high-energy neutrinos than for TeV and PeV 
$\gamma$-rays, that can be absorbed by
pair production with either the microwave background (CMB) or the infrared/optical photon background 
(IRB). Therefore the detection of astrophysical neutrinos can provide invaluable
complementary information on the existence of hadronic processes at astrophysical objects. 
Here we calculate the observable TeV-scale neutrino flux from RX~J1713.7-3946 and other sources.
This is not the first attempt to do such work. In fact, several calculations 
of the expected $\gamma$-ray and neutrino spectra were made for parametrized 
hadronic particles in the energy range of neutrino detectors \cite{AlvarezMuniz02,Costantini05,Vissani06,Lipari06,Kelner06,Kappes07,Beacom07}. 
In contrast to the earlier works, we include all relevant channels of neutrino production and 
also we do not rely
on assuming a simple power-law proton spectrum or a power-law with an exponential cutoff, because 
our neutrino production matrix can be applied to energetic particles with arbitrary spectrum.
\section{Inelastic cosmic-Ray interactions}
\subsection{Set-up of the Monte-Carlo event generator DPMJET-III}\label{Sec:HadronicReaction}
This work uses the same method to determine the $\gamma$-ray production matrix for
the cosmic-ray hadronic interactions as published earlier for $\gamma$-ray production \cite{Huang07}, 
namely applying the Monte-Carlo event generator, DPMJET-III \cite{Roesler00}, to simulate 
the secondary production in cosmic-ray interactions. The readers are referred to reference 
\cite{Huang07} for the details of this technique.
\subsection{Decay channels of secondary particles to leptons}\label{Sec:DecayChannel}
All secondary particles produced in simulated hadronic interactions are recorded while 
running the event generator DPMJET-III. The following decay modes show all the decay 
processes of secondary products with leptons as the final decay particles,
that are taken into account. Stable leptons considered in this work are 
$e^\pm,~\nu_e,~\bar{\nu}_e,~\nu_{\mu},~{\textrm{and}}~\bar{\nu}_{\mu}$. Note that the lifetime of 
neutrons is about 886 s, much shorter than the propagation time scale 
of cosmic-ray particles in the Galaxy. Therefore, the neutrons are treated as having decayed entirely in 
this work. The decay channels considered in 
this work are as follows. 
\begin{itemize}
\item baryonic decays:
\begin{itemize}
\item[] $n                  \rightarrow          p             +      e^-           +             \bar{\nu}_e$,
\item[] $\bar{n}            \rightarrow          \bar{p}       +      e^+           +             \nu_e$,
\item[] $\Lambda            \rightarrow $ $\left\{\begin{array}{l}
                                                   p    +      \pi^-,\\
                                                   n    +      \pi^0,
                                                  \end{array}
                                           \right.$
\item[] $\bar{\Lambda}      \rightarrow $ $\left\{\begin{array}{l}       
                                                  \bar{p}      +            \pi^+,\\
                                                  \bar{n}      +            \pi^0,
                                                  \end{array}
                                           \right.$
\item[] $\Sigma^0           \rightarrow          \Lambda       +      \gamma$,
\item[] $\Sigma^+           \rightarrow $ $\left\{\begin{array}{l}
                                                   p           +      \pi^0,\\
                                                   n           +      \pi^+,
                                                  \end{array}
                                           \right.$
\item[] $\Sigma^-            \rightarrow         n             +      \pi^-$.
\end{itemize}
\end{itemize}
\begin{itemize}
\item mesonic decays:
\begin{itemize}
\item[] $\pi^+              \rightarrow          \mu^+         +             \nu_{\mu}$,
\item[] $\pi^-              \rightarrow          \mu^-         +             \bar{\nu}_{\mu}$,
\item[] $\pi^0              \rightarrow          e^+           +             e^-           +      \gamma$,
\item[] $K^+                \rightarrow $ $\left\{\begin{array}{l}
                                                  \mu^+        +             \nu_{\mu},\\
                                                  \pi^+        +             \pi^0,
                                                  \end{array}
                                           \right.$
\item[] $K^-                \rightarrow $ $\left\{\begin{array}{l}
                                                  \mu^-        +             \bar{\nu}_{\mu},\\
                                                  \pi^-        +             \pi^0,
                                                  \end{array}
                                           \right.$
\item[] $K^0_S              \rightarrow $ $\left\{\begin{array}{l}
                                                  2\pi^0,\\
                                                  \pi^-        +             \pi^+,
                                                  \end{array}
                                           \right.$
\item[] $K^0_L              \rightarrow $ $\left\{\begin{array}{l}
                                                  3\pi^0,\\
                                                  \pi^-        +             \pi^+         +             \pi^0,\\
                                                  \pi^+        +             e^-           +             \bar{\nu}_e,\\
                                                  \pi^-        +             e^+           +             \nu_e,\\
                                                  \pi^+        +             \mu^-         +             \bar{\nu}_\mu,\\
                                                  \pi^-        +             \mu^+         +             \nu_\mu.\\
                                                  \end{array}
                                           \right.$
\end{itemize}
\end{itemize}
\begin{itemize}
\item leptonic decays:
\begin{itemize}
\item[] $\mu^+              \rightarrow          e^+           +            \nu_e          +      \bar{\nu}_\mu$,
\item[] $\mu^-              \rightarrow          e^-           +            \bar{\nu}_e    +      \nu_\mu$,
 \item[] $\tau^+             \rightarrow          \left\{\begin{array}{l}
                                                 \mu^+         +            \nu_\mu        +      \bar{\nu}_\tau, \\
                                                 e^+           +            \nu_e          +      \bar{\nu}_\tau ,
                                                  \end{array}
                                           \right.$

\item[] $\tau^-             \rightarrow          \left\{\begin{array}{l}
                                                 \mu^-         +            \bar{\nu}_\mu  +      \nu_\tau, \\
                                                 e^-           +            \bar{\nu}_e    +      \nu_\tau ,
                                                  \end{array}
                                           \right.$
%
%
%
\end{itemize}
\end{itemize}

In the DPMJET simulation setup, the decay of $\eta$ mesons produced in the hadronic interactions has already been treated in the simulation. In our study on cosmic-ray 
induced $\gamma$-ray production \cite{Huang07}, we have verified that the $\eta$ mesons are properly accounted for by performing test runs under different PYTHIA 
parameters, for which the $\eta$ decays can be avoided in the simulation. We have also performed another independent verification using the very small 
contribution to $e^-/e^+$ and $\mu^-/\mu^+$ pairs from $\eta$ decays, which are found consistent within the statistical uncertainty with all $\eta$ mesons having decayed.

The $\pi^\pm$ decays and subsequent $\mu^\pm$ decays via the $\pi-\mu-e$ decay chain
are dominant in cosmic-ray interactions because pions carry the highest
multiplicity among all the secondaries. We ignore neutrino production by charmed particles such 
as $D$ and $\bar{D}$, because they contribute only at $E_\nu \gtrsim$ 100 TeV \cite{Honda95}, where 
they are ignorable compared with neutrino production by pions.
In the case of decays of moving $\pi^\pm$'s, the $\pi$ energy is roughly equally divided among the four decay products, and thus the flux ratios are approximately 
$(\nu_e +\bar{\nu}_e)/(\nu_\mu+\bar{\nu}_\mu) =1/2$ and $\bar{\nu}_\mu/\nu_\mu =1$, 
no matter what the $\pi$ spectrum is. In reality, however, these neutrino ratios may vary on account of the decays of heavier baryons and mesons. An excess 
of $\mu^+$ multiplicity arises because more $\pi^+$'s than $\pi^-$'s are produced, and so the ratios 
$\nu_e /\bar{\nu}_e$ and $\bar{\nu}_\mu/\nu_\mu$ are somewhat increased.

For the two-body decay processes, the decay spectra are evaluated by particle kinematics; 
for the electrons/positrons from muon decays, the decay spectra are calculated by 
Lorentz transformation of the particle distribution in the center-of-mass system of 
the muon \cite{Scanlon65}, which also includes the effect of the $\mu$ polarization \cite{Volkova88,Barr88}; 
for kaon decays and heavy nucleon decays, the Dalitz-plot
distribution is applied. The values published in Particle Data Group are employed 
for the fraction of each individual decay process. 
\subsection{Resonance contribution at low energies}
In our earlier work on $\gamma$-ray production \cite{Huang07} we already found that DPMJET is unreliable below a few GeV
collision energy. At these energies we therefore use a parametric model \cite{Kamae06} to calculate the 
resultant decay spectra of $\gamma$, $e^\pm$, $\nu_e$ ($\bar{\nu}_e$) and $\nu_{\mu}$ ($\bar{\nu}_{\mu}$). 
For the decay of the resonances $\Delta(1232)$ and $\Delta(1600)$ we assume
the pion momentum to be isotropically distributed in the center-of-mass system
and no angular correlation between the two pions in $\Delta(1600)$ decay. 

Note that the parametric model for pion and resonance production at low incident energies is
for pp interactions only. We then use DPMJET to calculate the 
energy-dependent weight factors, which allow us to parametrically account for p+ISM and He+ISM 
collisions in the parametrization approach, even though it is derived for pp collisions only. 
The weight factors carry no strong dependence on the energy of the projectile particle.
\section{The Production Matrix of Secondary Leptons Generated by Cosmic Rays}\label{Section:ProductionMatrix}
The differential production rate of a final secondary particle is given by
\begin{eqnarray}\label{EQ:2ndSpectra}
Q_{\textrm{2nd}}(E) 
= \frac{dn}{dt\cdot dE \cdot dV}
=n_{ISM} \int_{E_{CR}} dE_{CR}\, N_{CR}(E_{CR})\,c\beta_{CR}\left(\sigma \frac{dn}{dE}\right)
\end{eqnarray}
with $N_{CR}(E_{CR}) =\frac{dn_{CR}}{dE_{CR}\cdot dV} ~\textrm{(GeV~cm}^3)^{-1}$ as the 
differential density of CR particles (p or $\alpha$). The differential cross section of a secondary 
particle produced in (p,$\alpha$)-ISM collisions is 
\begin{eqnarray}\label{EQ:ProdXSection}
\frac{d\sigma}{dE}(E_{CR},E) = \sigma_{prod} \frac{dn}{dE} 
\end{eqnarray}
where $\sigma_{prod}$ is the inelastic production cross section, i.e., the sum of diffraction, 
non-diffraction and resonance components whose values are calculated by DPMJET 
and the parametric model; $\frac{dn}{dE}$ is the multiplicity spectrum of the secondary 
particle in question.

We follow the decay of unstable secondary particles to stable particles, and thus
obtain the final spectrum of the stable secondaries $\gamma$-rays, $e^\pm$, $\nu_e$, $\bar{\nu}_e$, 
$\nu_\mu$ and $\bar{\nu}_\mu$ as 
\begin{equation}\label{EQ:Final2ndSpectra}
Q_{\textrm{2nd}}(E_i) 
= \sum_k n_{ISM}\int_{E_{CR}} dE_{CR}\ N_{CR}(E_{CR})\, c\beta_{CR}\, \sigma (E_{CR})\,
\frac{dn_{k,i}}{dE_i}(E_i, E_{CR})
\end{equation}
where $\frac{dn_{k,i}}{dE_i}$ is the multiplicity spectrum of stable particle species $i$ 
resulting from production channel $k$, either an unstable secondary or direct production. 
For binned particle spectra the production integral, Eq.~(\ref{EQ:Final2ndSpectra}), can be re-written as
\begin{eqnarray}
Q_{\textrm{2nd}}(E_i) 
&=& n_{ISM} \sum_{j} \, \Delta E_{j}\ N_{CR}(E_j)\,c\beta_j\,\sigma (E_{j})\,\sum_{k}\,\frac{dn_{k,i}}{dE_i}(E_i,E_j) \label{EQ:EBinning}\\
&=& n_{ISM} \sum_j   \,\Delta E_{j}\, N_{CR}(E_j)\, c\beta_{j}\, \sigma_j \, \mathbb{M}_{ij}  \label{EQ:ProductionMatrix}
\end{eqnarray}
for the secondary particle of interest. In Eqs. (\ref{EQ:EBinning}) and (\ref{EQ:ProductionMatrix}), the energies for cosmic-ray particles ($p$ or $\alpha$) and 
the secondary particles are parametrized as \cite{Huang07}:
\begin{eqnarray}
E_T &=& 1.24\cdot (1+0.05)^j         \hspace{39pt}\textrm{GeV/n,~for cosmic-rays,} \label{EQ:ECR}\\
E_k &=& 0.01\cdot (1.121376)^{i-0.5} \hspace{20pt}\textrm{GeV, for secondary particles.}\label{EQ:E2nd}
\end{eqnarray}
$n_{ISM}\simeq 1.11~n_H$ in Eq.~(\ref{EQ:ProductionMatrix}) accounts for the specific element abundance in the ISM assumed 90\% H, 10\% He, 0.04\% O and 0.02\% C here. 
Please note that the ratio of the $\gamma$-ray and neutrino yields is indepedendent of $n_{ISM}$. The main product of our work is the production matrix $\mathbb{M}_{ij}$
that describes the production spectrum of the stable secondary particles, one for each, 
for arbitrary cosmic-ray spectra, separately for protons and Helium nuclei. 
\section{Applications: SNR}
With the production matrix established for each stable particle, we are now in the position to calculate
the production spectra of secondary leptons and neutrinos in a variety of sources. As an example we here
consider young shell-type SNRs. Being associated with dense gas along the line-of sight, the shell-type SNR RX~J1713.7-3946 could provide 
significant information on the acceleration of hadronic cosmic-rays. 
In addition, RX J1713.7-3946 is also a unique supernova remnant in the 
sense that its X-ray emission is strongly dominated by a non-thermal component, which is presumed to 
be synchrotron radiation of ultrarelativistic electrons. These electrons also
produce TeV-band gamma-rays, but would require a relatively small magnetic field to account for
the TeV-band $\gamma$-ray spectrum observed with HESS \cite{Aharonian06RXJ1713.7-3946}. 
The recently observed rapid variability of X-rays of RX J1713.7-3946 has been interpreted
as evidence of a multi-mG magnetic field \cite{Uchiyama07}, in which case the leptonic scenario 
would be untenable.

We can test this interpretation by calculating the synchrotron radiation of secondary electrons in
this remnant. In an earlier paper we have determined the cosmic-ray spectrum in RX~J1713.7-3946
under the assumption that the observed TeV-band spectrum is hadronic in origin \cite{Huang07}.
Using our newly derived production matrix for secondary electrons and positrons, we can determine
the differential production rate of those particles which, together with a time-dependent
cosmic-ray continuity equation, allows us to calculate the radio synchrotron radiation of the 
secondary electrons with mainly the magnetic field strength as a fully free parameter.

Our fit of the TeV $\gamma$-ray spectrum constrains only the strongly curved cosmic-ray spectrum in
the 1-100~TeV energy range. However, synchrotron emission in the GHz range is produced by
GeV-scale electrons and we do not know how the spectrum continues down to GeV energies. So for 
simplicity, we match a single-power-law spectrum to the multi-TeV spectrum of cosmic-ray nucleons that we derive from 
fitting the HESS data,
\begin{eqnarray}
&N(E)=& N_0\,\left({E\over {E_0}}\right)^{-s+\sigma\,\ln {E\over {E_0}}}
\exp\left( -\frac{E}{E_{\textrm{max}}}\right ),  \quad  E \ge E_c \label{EQ:SpectraRXJ1713.7-3946}\\
&N(E)=& M_0 \left({E\over {E_0}}\right)^{-\alpha},  \hspace*{120pt}  E < E_c \label{EQ:CRSpectraLowE}
\end{eqnarray}
where the parameters in equation~(\ref{EQ:SpectraRXJ1713.7-3946}) are derived by fitting 
the observed TeV-band $\gamma$-ray spectrum of RX J1713.7-3946 \cite{Huang07,Huang07ICRC}.
In this parametrization, $E_0=15$~TeV is a normalization 
chosen to render uncorrelated the variations in the power-law index, $s$, and the spectral 
curvature, $\sigma$. $E_{\rm max}$ is the cut-off energy, and the best-fitting parameter values 
are $s=2.13$, $\sigma=-0.25$, and $E_{\rm max}\gtrsim 200$~TeV. $N_0$ is the overall normalization, 
determined by the integral spectrum of $\gamma$-rays above 1 TeV 
\cite{Aharonian06RXJ1713.7-3946}. If a numerical value of $\alpha$ is set, the normalization $M_0$ and the merging energy 
$E_c$ are fully determined by the two continuity conditions
that link the two spectral forms (\ref{EQ:SpectraRXJ1713.7-3946}) and (\ref{EQ:CRSpectraLowE}). 
Models of particle acceleration in cosmic-ray modified shocks 
suggest that the spectral index $\alpha$ should be slightly smaller than 2 \cite{Amato06,Berezhko99}. 
Note that $\alpha$ denotes the average particle spectral index between about 1~GeV and 10~TeV.
The combined cosmic-ray spectrum together with our production matrices for electrons and 
positrons then yields the differential production rate of secondary leptons that is implied 
by the $\gamma$-ray spectrum observed from RX~J1713.7-3946.
\subsection{Electrons/Positrons}
Secondary electrons have been accumulated in the remnant since it commenced hadron acceleration.
To determine the synchrotron radiation of the secondary electrons, we must
obtain the electron spectrum in the SNR by solving a simplified time-dependent electron transport equation:
\begin{eqnarray}\label{EQ:ElectronTransport}
\frac{\partial N}{\partial t} + \frac{\partial}{\partial E}\left( b(E) N \right) = Q(E,T)
\end{eqnarray}
where $T$ is the time after supernova explosion and
the electron energy loss rate is that for synchrotron radiation 
\begin{eqnarray}\label{EQ:SynchrotronLoss}
b(E)=\dot{E}= -\frac{4}{3}\,\sigma_{\textrm{Th}}\,c\, U_B\,\beta^2\, \gamma^2 
=-\frac{4}{3}\,\sigma_{\textrm{Th}}\,c\,\frac{B^2}{2\mu_0}\,\beta^2\, \gamma^2  
\end{eqnarray}
with $U_B$ as the energy density of the magnetic field and $\sigma_{\textrm{Th}}$ as the Thomson cross section.

The likely distance to RX~J1713.7-3946 is 1~kpc and the implied age would be about 1,600 years
\cite{Koyama97,Uchiyama07,Wang97}. An alternative kinematic solution places the remnant at about 6~kpc
in distance with implied age of 10,000~years \cite{Slane99}. Whatever the acceleration history of the remnant is,
cosmic-ray particles which have been accelerated early in its evolution will have lost some of their energy 
by adiabatic expansion, with most of it happening very early in the evolution. To be conservative
we therefore consider only half of the remnant age, during which we assume the differential 
production rate of secondary electrons, $q(E)$, to be constant and given by our production matrix
and the cosmic-ray spectrum as in Eqs.~(\ref{EQ:SpectraRXJ1713.7-3946}) and (\ref{EQ:CRSpectraLowE})
\begin{eqnarray}\label{EQ:ElectronSource}
Q(E,T) = q(E) \, \Theta(T-t_{1/2})\ .
\end{eqnarray}
Then the spectrum of secondary electrons or positrons is obtained by integrating 
the Green function and the source term over time and energy:
\begin{eqnarray}\label{EQ:ElectronPopulation}
N(E,T) =\int \int dE_0\, dT_0\ q(E_0)\,\Theta(T_0-t_{1/2})\,\frac{\delta(T-T_0-\tau)}{|b(E)|}\, 
\Theta\left(E_0-E\right)
\end{eqnarray}
where $\tau=\int_{E_0}^E \frac{dE'}{b(E')}$ is the electron cooling time. For high energies 
$\tau \le t _{1/2}$ and the spectrum is 
loss-limited, i.e. essentially steepened by 1 in the spectral index, whereas at low energies the 
production spectrum is preserved. Note that the synchrotron flux calculated using the electron spectrum
(\ref{EQ:ElectronPopulation}) is absolutely normalized by the TeV $\gamma$-ray flux.

\begin{figure}
\begin{center}
\includegraphics[scale=0.65]{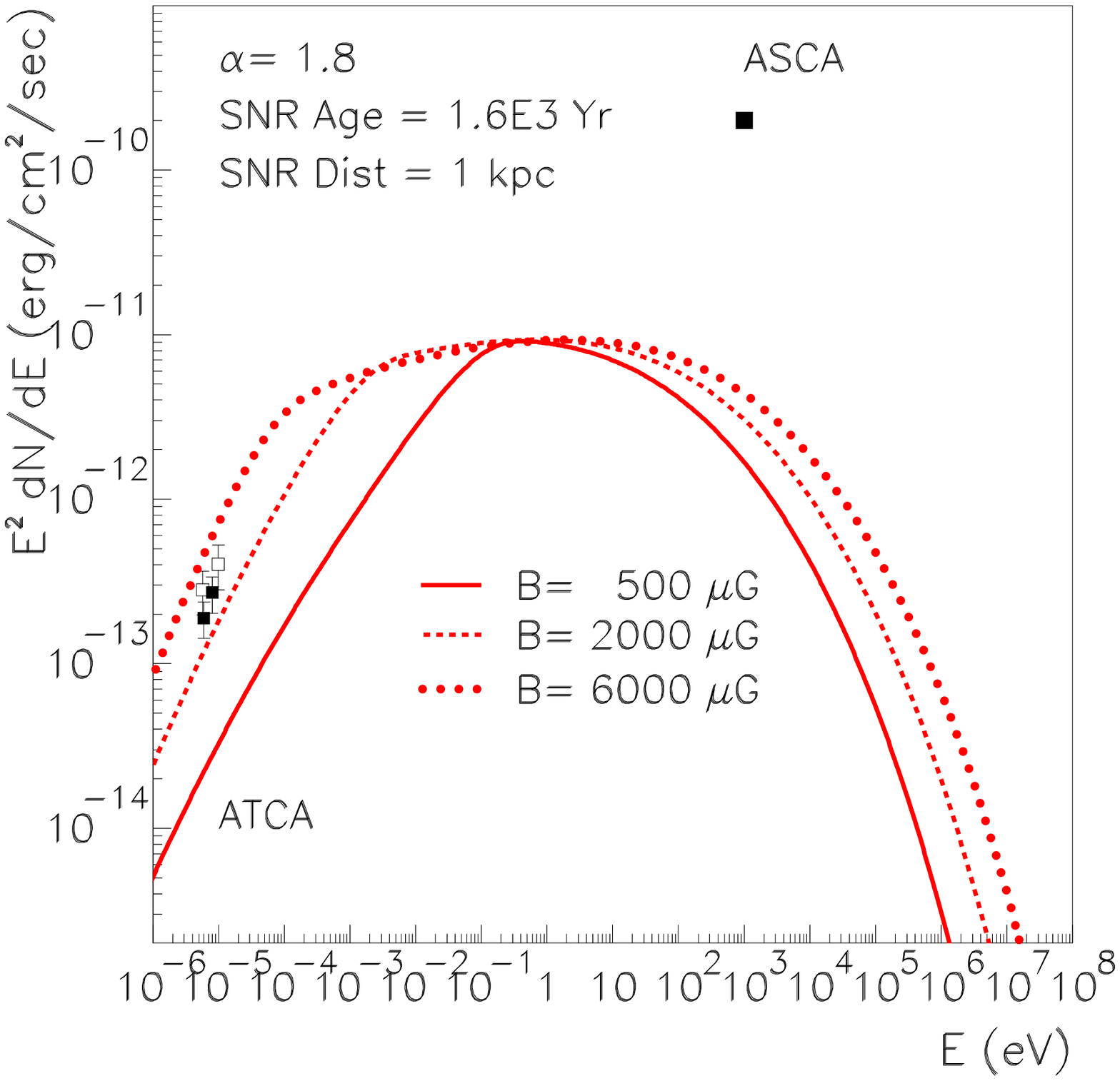}
\caption{The synchrotron radiation contributed by secondary electrons produced in RX~J1713.7-3946
assuming a source distance of 1 kpc and the SNR age as 1,600 years. The cosmic-ray particle spectrum 
follows (\ref{EQ:SpectraRXJ1713.7-3946}) 
and (\ref{EQ:CRSpectraLowE}) with $\alpha=1.8$. The curves show the
synchrotron spectra for magnetic fields $B = 500,~ 2000,~\textrm{and}~6000$ $\mu$G. 
The ATCA data (open marks) are based the brightness of the NW
rim of RX J1713.7-3946 \cite{Lazendic04} scaled to the entire remnant with a factor 3 based on
the angular distribution of TeV gamma-ray emission \cite{Aharonian06RXJ1713.7-3946}. The data points (closed marks) are taken
from \cite{Aharonian06RXJ1713.7-3946}, where they are referenced as private communication. See text for discussions.}
\label{Fig:SynchSpectraCR180}
\end{center}
\end{figure}

Figure \ref{Fig:SynchSpectraCR180} shows an example of the synchrotron spectrum from RX~J1713.7-3946
that is contributed by secondary electrons assuming a distance of 1 kpc,
the age as 1,600 years, and the power-law index $\alpha=1.8$ for the multi-GeV cosmic-ray nucleon
spectrum. The synchrotron spectra for three different values of the
magnetic fields, $B= 500,~ 2000,~\textrm{and}~6000$ $\mu$G, are shown in comparison with the
observed X-ray \cite{Hiraga05} and radio \cite{Lazendic04,Aharonian06RXJ1713.7-3946} data. 
At higher frequencies the synchrotron power is always about half
the TeV-band $\gamma$-ray power on account of the equal likelihood of neutral and charged pions 
being produced in hadronic collisions. The non-thermal X-ray emission is therefore always 
contributed by primary electrons.

To be noted from the figure is that for a field strength of 2~mG the radio synchrotron flux is similar
to the observed flux, so the primary electrons must be very few because they would contribute only the 
remaining fraction of the observed radio flux, thus further lowering
the $e/p$ ratio in accelerated particles \cite{Katz07}. This limit is much more severe, if 
the distance to RX~J1713.7-3946 is more than 1~kpc, and the age correspondingly more than 1,600~years. 
It is also more severe, if the effective GeV-to-TeV spectral index of cosmic-ray nucleons in the 
source is softer, $\alpha\ge 1.8$, and it is relaxed in the opposite case. One should note 
that in mG-fields GHz-band synchrotron emission is 
emitted by electrons with about 300~MeV kinetic energy, i.e. below the relativistic transition. 
The effective spectral index must be measured by comparing the
$1$~GeV flux and the $\sim 10$~TeV flux of cosmic-ray nucleons, which is considerably softer than
the local spectral index near 1~TeV. Measurements of the radio synchrotron of a large number of 
shell-type SNRs indicate that in the GeV band, the spectral index is locally around 2 \cite{Green01}, 
whereas at 1-10~TeV the local
spectral index should be closer to 1.5 \cite{Amato06}, so on average we may expect $\alpha\simeq 1.8$.
Consequently, a multi-mG
magnetic field is unlikely to exist in a large fraction of the synchrotron emission region
or $\alpha$ must be significantly smaller than 1.8 (i.e., the spectrum is harder).
This result is conservative, because we are ignoring secondary electrons that were produced in the 
first half of the time since the SN explosion.

The recent observation of rapid variability of X-rays of RX J1713.7-3946 \cite{Uchiyama07} has been interpreted as evidence for the notion that  the magnetic field of 
the hot spots (two compact regions) could be as large as mG level. These compact regions are in only a few arcseconds and likely present transient acceleration sites of
high-energy electrons. Our calculations shows that, if  such strong magnetic field exists in SNR shell, they are unlikely to fill the entire remnants.

Table 1 summarizes the limits on the averaged magnetic-field strength for various values of the effective 
GeV-to-TeV spectral index of cosmic-ray nucleons and different values for the age of RX~J1713.7-3946. 
The contribution of synchrotron radiation by secondary electrons is compared with the integrated flux density 20 Jy at 1.4 GHz, 
which is the result of \cite{Lazendic04} for the NW region but scaled to the entire remnant with a factor 3 based on
the angular distribution of TeV gamma-ray emission \cite{Aharonian06RXJ1713.7-3946}. This scaling factor carries a sizable uncertainty not less than 30\%, largely owing to
systematic uncertainty in the background subtraction in the radio data. Some similar works \cite{Aharonian06RXJ1713.7-3946,Slane99,Lazendic04} have used a scaling factor 
of 2 for the entire remnant. 
\subsection{Neutrinos}
\begin{figure}
\begin{center}
\includegraphics[scale=0.65]{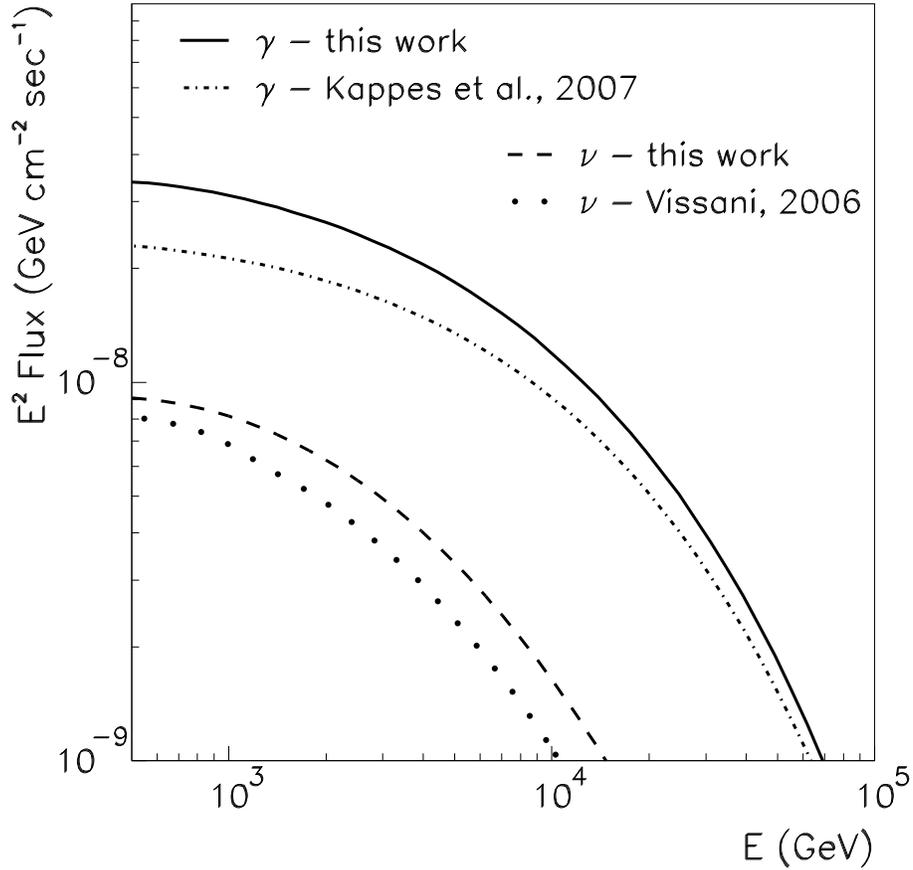}
\caption{The $\gamma$-ray and raw $\nu_\mu$ spectra from the cosmic-ray source 
RX J1713.7-3946 derived using the production matrix discussed in this work, shown in
comparison with results of published parametrizations \cite{Kelner06,Kappes07,Vissani06} based on the
same cosmic-ray particle spectrum. With the same generating cosmic-ray particle spectrum, our production matrix gives about 15-20\% more 
$\gamma$-rays at all energies. Our calculation also shows about 30\% more neutrinos than the parametric model of Kappes et al. \cite{Kappes07} after considering all 
relevant decay processes in the cosmic-ray hadronic interactions. The calculations on $\nu_\mu$ spectra indicate that the parametric model of 
Vissani \cite{Vissani06} gives an overall lower raw $\nu_\mu$ yield for the same $\gamma$-ray flux before taking neutrino oscillations into account.}
\label{Fig:ESpectraGammaMuNuRXJ1713.7-3946}
\end{center}
\end{figure}

High-energy cosmic neutrinos are inevitably produced in parallel with hadronic
$\gamma$-rays. Therefore, TeV-band $\gamma$-ray
sources represent the prime targets to search for cosmological neutrinos. 
Several calculations on the expected $\gamma$-ray and neutrino spectra were 
made for parametrized spectra of hadronic particles in the energy range of neutrino 
detectors, $E\approx ~(1-1000)~\textrm{TeV}$ \cite{AlvarezMuniz02,Costantini05,Vissani06,Lipari06,Kelner06,Kappes07,Beacom07}, usually
by assuming a simple power-law proton spectrum or a power-law 
with an exponential cutoff. In contrast, our neutrino production matrix 
can be applied to cosmic-ray nucleon spectra of any form. Also,
the particle yields in the parameterizations can be underestimated because they are often based only on the pion channels.
Our production matrix accounts for all relevant decay processes in cosmic-ray hadronic interactions. As shown in 
Figure~\ref{Fig:ESpectraGammaMuNuRXJ1713.7-3946}, we find that using the same cosmic-ray spectrum, our production matrix gives about 15-20\% more $\gamma$-rays at 
all energies. Our production matrix also yields an approximately 30\% more muon neutrinos than the parametric model \cite{Kappes07} after full mixing. 
The excess of $\gamma$-rays and neutrinos in this figure agrees with the conclusion of our earlier study \cite{Huang07}, which has considered the full picture of 
$\gamma$-ray production in hadronic interactions and indicated about 20\% more hadronic $\gamma$-rays relative to $\gamma$-rays from $\pi^0$ decays alone. 
Also note that $\eta$-decays contribute about 8-10\% and kaons also have roughly the same contribution \cite{Huang07}. 
Only a few percent of $\gamma$-rays are contributed by other decay processes (together with the direct $\gamma$-ray production). 
For a given $\gamma$-ray or neutrino flux, our production matrix therefore implies a lower flux of cosmic-ray nucleons than estimated using the parametric 
models \cite{Kelner06,Kappes07}.

In high-energy hadronic interactions, the resulting initial $\nu$ flavor ratio from $\pi$ decays is $\nu_e:\nu_\mu:\nu_\tau=1:2:0.$ However, the neutrino 
oscillations transform this ratio to 1:1:1, i.e., full mixing, because the SNR size of $\sim$ 10 pc implies a substantial range of neutrino path lengths, wider than 
the neutrino oscillation wavelength 
$L_{\textrm{OSC}} = (4\pi E/\Delta m^2)\hbar c \simeq 0.8~(E/\textrm{TeV}) (\Delta m^2/10^{-10}\textrm{eV}^2)^{-1}~\textrm{pc}$ \cite{Crocker02}. Therefore, 
it is reasonable to expect the $\nu_\mu$ spectrum 
arriving at Earth to be about 1/3 of the total $dN_\nu/dE$. The resulting $\nu_\mu$ spectrum is only weakly dependent of the relative fraction of $\nu_e$'s 
and $\nu_\mu$'s at the source. Other studies \cite{Costantini05,Vissani06} have calculated high-energy neutrinos from galactic sources considering the actual 
neutrino mixing angles for point sources. For a comparison of the neutrino source functions by our production matrix and the parametric model \cite{Vissani06}, 
we therefore must consider the $\nu_\mu$ spectra from the TeV $\gamma$-ray source RX~J1713.7-3946 before neutrino oscillations. 
In Figure \ref{Fig:ESpectraGammaMuNuRXJ1713.7-3946}, the $\nu_\mu$ spectrum is calculated using the production matrix and a generating cosmic-ray particle spectrum
\cite{Huang07} that is determined as best-fitting the observed HESS $\gamma$-ray data of RX~J1713.7-3946; the $\nu_\mu$ spectrum by the parametric model \cite{Vissani06} 
is calculated based on the same $\gamma$-ray spectrum. As seen in this figure, the parametric model gives overall lower raw $\nu_\mu$ yield from RX~J1713.7-3946.

\begin{figure}
\begin{center}
\includegraphics[scale=0.65]{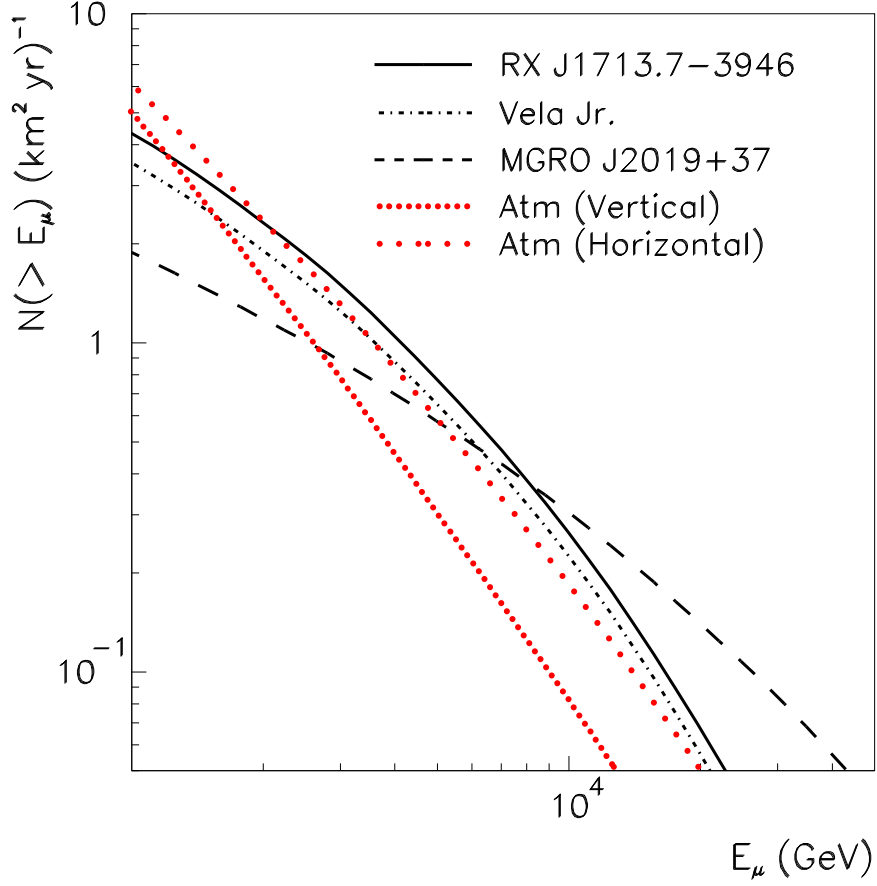}
\caption{Integrated fully mixed ($\nu_\mu + \bar{\nu}_\mu$)-induced muon rates from the TeV $\gamma$-ray 
sources RX~J1713.7-3946 (solid line), 
Vela Jr. (dash-dot line) and MGRO J2019+37 (dash line) above a given muon energy within an IceCube-like detector. 
The generating cosmic-ray spectra are obtained by fitting the observed TeV $\gamma$-rays of RX J1713.7-3946 
\cite{Aharonian06RXJ1713.7-3946}, 
Vela Jr. \cite{Aharonian07VelaJr}, and MGRO J2019+37 \cite{Kistler06}, using the production matrices for 
gamma-rays and neutrinos. 
The atmospheric background rates, vertical (long dot) and horizontal (short dot), are calculated using the
parametrization of \cite{Volkova80} and assuming an angular resolution element of 3~deg$^2$.}
\label{Fig:IceCubeMuMuNu}
\end{center}
\end{figure}

Practically, neutrinos and anti-neutrinos are not distinguishable in neutrino telescopes. Therefore, full neutrino mixing is usually assumed for the $\nu$ 
spectrum. Figure \ref{Fig:IceCubeMuMuNu} shows the $\nu_\mu$-induced muons which may be detected 
in an IceCube-like experiment \cite{Beacom07} from TeV $\gamma$-rays sources such as  
RX~J1713.7-3649 \cite{Aharonian06RXJ1713.7-3946}, MGRO J2019+37 \cite{Beacom07} and Vela Jr. \cite{Aharonian07VelaJr}, 
together with the background from atmospheric neutrinos. The $\nu_\mu$-induced muons for each source example and the atmospheric background are calculated 
by a parametric model \cite{Kistler06}, including the neutrino attenuation due to scatterings within the Earth.

In calculating the $\nu_\mu$-induced muon event rate, the generating cosmic-ray particle spectrum for each source was obtained by fitting the observed $\gamma$-ray 
spectra using our gamma-ray production matrix \cite{Huang07}, from which the $\nu$-spectra are determined using the neutrino production matrix (\ref{EQ:ProductionMatrix}) 
and assuming full mixing. Note that a TeV $\gamma$-ray spectrum from MGRO J2019+37 is not available at this time. Therefore a fit was performed using EGRET 
data for 3EG J2021+3716 \cite{Beacom07}. A more reliable calculation will depend on a measurement of the TeV $\gamma$-ray spectrum of this source.

The charged-current and neutral-current cross sections, the inelasticity involved in $\nu N \rightarrow \mu^\pm +X$ interactions, and the range of the muons in the 
detector are based on the theoretical results of \cite{Kistler06,Gandhi96,Gandhi98}. The $\nu$-induced muon event rate for the
atmospheric background is then calculated using the parametrized atmospheric neutrino background 
as derived in \cite{Volkova80}, both vertically and horizontally, and assuming an angular resolution element of 3~deg$^2$ (1 deg uncertainty radius), which roughly 
corresponds to the angular resolution of IceCube. As seen in Figure \ref{Fig:IceCubeMuMuNu}, the calculations indicate a good detectability of high-energy neutrinos from 
the high-energy $\gamma$-ray sources at a few TeV in muon energy, or about 10~TeV neutrino energy. An accumulation of 
data over about five to ten years would allow testing the hadronic origin of TeV $\gamma$-rays.


The lepton and neutrino production matrices are applicable for arbitrary cosmic-ray hadron spectra. The main systematic uncertainty is that of the simulation tool 
DPMJET itself. DPMJET has been extensively tested against data for accelerator experiments \cite{Huang07,Roesler00,Ranft95}, for the simulation of cosmic-ray air 
shower problems \cite{Knapp03,Antoni01}, and for the atmospheric $\gamma$-ray and neutrino
production \cite{Kasahara02}. Although the uncertainty for hadronic productions simulated by DPMJET differs with simulation energy as well as the interaction products, it
is as good as at least within 15\% \cite{Ranft95,Antoni01}.

\section{Conclusion}
In addition to Galactic $\gamma$-ray emission, the Galactic neutrino emission could 
provide complementary information on the existence of hadronic
processes in astrophysical objects and also on the origin of Galactic cosmic-rays. 
Here we have calculated the $e^\pm$, $\nu_e$ and $\nu_\mu$ production
in cosmic-ray interactions and derived matrices that allow to estimate the spectra of secondary leptons 
for arbitrary spectra of the parent hadronic cosmic rays.

With the production matrices, we calculate the synchrotron radiation of secondary electrons
in the shell-type SNR RX~J1713.7-3649 under the assumption that the observed TeV-scale gamma-ray emission
is hadronic in origin. We find that the radio synchrotron flux of the secondary electrons exceeds 
the observed flux level if the magnetic field strength is too high. The multi-mG fields recently invoked
to explain the X-ray flux variations \cite{Uchiyama07} are therefore unlikely to extend over a large 
fraction of the radio-emitting region or the spectrum of hadronic cosmic rays in the 0.1-100~GeV energy
window must be unusually hard.

Our calculations for neutrinos show that the neutrino detection rates from 
the TeV $\gamma$-ray sources RX J1713.7-3946 and Vela Jr. are promising for muon energies of a few TeV,
or in other words, about 10~TeV neutrino energy. An accumulation of 
data over about five to ten years would allow testing the hadronic origin of TeV $\gamma$-rays.

The production matrices for $e^-$, $e^+$, $\nu_e$, $\bar{\nu}_e$, $\nu_\mu$ and $\bar{\nu}_\mu$ will be 
made available for download at {\em http://cherenkov.physics.iastate.edu/lepton-prod}.
\acknowledgments
The author C.-Y. Huang would like to thank Y. Liu for his helpful comments. Support by NASA under award No. NAG5-13559 is gratefully acknowledged.

%
%
\clearpage

\clearpage

\clearpage

%
%
\clearpage
\begin{table}
\caption{Relative contribution of synchrotron radiation by secondary electrons in 
RX~J1713.7-3946 to the observed radio flux in different source models and magnetic fields
near the SNR shell. The relative contribution of synchrotron radiation in this table are calculated 
relative to the radio data of integrated flux density 20 Jy at 1.4 GHz, 
which is the result of \cite{Lazendic04} scaled to the entire remnant with a factor 3 based on
the angular distribution of TeV gamma-ray emission \cite{Aharonian06RXJ1713.7-3946}.} \label{Table:SynchSpectraVsB}
\vspace*{0.25cm}
\begin{tabular}{|c|c|cr|cr|cr|cr|}
\hline
index \quad   & Age (Yr)  & B ($\mu$G) & ratio\quad   & B ($\mu$G) & ratio\quad  & B ($\mu$G) & ratio\quad  & B ($\mu$G) & ratio\quad    \\ \hline
$\alpha=1.6$  & 1,600     & 500        & 3.4\%        & 1000       &   8\%       & 2000       &  17\%       & 4000       &  40\% \\
$\alpha=1.8$  & 1,600     & 500        &   8\%        & 1000       &  18\%       & 2000       &  41\%       & 4000       &  95\% \\
$\alpha=1.8$  & 10,000    & 500        &  48\%        & 1000       & 111\%       & 2000       & 255\%       & 4000       & 594\% \\
$\alpha=2.0$  & 1,600     & 500        &  21\%        & 1000       &  50\%       & 2000       & 119\%       & 4000       & 279\% \\
\hline            
\end{tabular}
\end{table}


\clearpage
\begin{thebibliography}{99}
\bibitem{Huang07}
C.-Y. Huang, S.-E. Park, M. Pohl and C. D. Daniels, \AstroPartPhys{27}{2007}{429}
\bibitem{Roesler00}
S. Roesler, R. Engel and J. Ranft, 
Advanced Monte Carlo for Radiation Physics, Particle Transport Simulation and Applications
(MC 2000), Lisbon, 2000
\bibitem{Blattnig00} 
Steve R. Blattnig, Sudha R. Swaminathan, Adam T. Kruger, Moussa Ngom and John W. Norbury, \PRD{62}{2000}{094030}
\bibitem{Kamae06}
T. Kamae, N. Karlsson, M. Tsunefumi, A. Toshinori and K. Tatsumi, \APJ{647}{2006}{692}
\bibitem{Koyama95}
K. Koyama, R. Petry, E. V. Gotthelf et al., \Nature{378}{1995}{255}
\bibitem{Koyama97}
K. Koyama, K. Kinugasa, K. Matsuzaki et al., \PASJ{49}{1997}{L7}
\bibitem{Keohane97}
J. W. Keohane, R. Petry, E. V. Gotthelf et al., \APJ{484}{1997}{350}
\bibitem{Slane99}
P. Slane, B. M. Gaensler, T. M. Dae et al., \APJ{525}{1999}{357}
\bibitem{Allen97}
G. E. Allen, J. W. Keohane, E. V. Gotthelf et al., \APJ{487}{1997}{L97}
\bibitem{Borkowski01}
K. J. Borkowski, J. Rho, S. P. Reynolds and K. K. Dyer, \APJ{550}{2001}{334} 
\bibitem{Pohl96}
M. Pohl, \AA{307}{1996}{L57}
\bibitem{Aharonian05VelaJr}
The HESS Collaboration, F. A. Aharonian et al., \AA{437}{2005}{L7}
\bibitem{Aharonian06RXJ1713.7-3946}
The HESS Collaboration, F. A. Aharonian et al., \AA{449}{2006}{223}
\bibitem{Reimer02}
O. Reimer and M. Pohl, \AA{390}{2002}{L43}
\bibitem{Katz07}
B. Katz, E. Waxman, \JCAP{Accepted}{2008}{(arXiv:0706.3485)}
\bibitem{Hiraga05}
J. S. Hiraga, Y. Uchiyama, T. Takahashi and F. A. Aharonian, \AA{431}{2005}{953}
\bibitem{Drury01}
L. O'C. Drury et al., \SpaceSciRev{99}{2001}{329}
\bibitem{Uchiyama07}
Y. Uchiyama, F. A. Aharonian, T. Tanaka, T. Takahashi and Y. Maeda, \Nature{449}{2007}{576}
\bibitem{Takahashi07}
T. Takahashi et al., \PASJ{Accepted}{2008}{(arXiv:0708.2002)}
\bibitem{AlvarezMuniz02}
J. Alvarez-Mu\~{n}iz and F. Halzen, \APJ{576}{2002}{L33}
\bibitem{Costantini05}
M. L. Costantini and F. Vissani, \AstroPartPhys{23}{2005}{477}
\bibitem{Vissani06}
F. Vissani, \AstroPartPhys{26}{2006}{310}
\bibitem{Lipari06}
P. Lipari, \NIMA{567}{2006}{405}
\bibitem{Kelner06}
S. R. Kelner, F. A. Aharonian and V. V. Bugayov, \PRD{74}{2006}{034018}
\bibitem{Kappes07}
A. Kappes, J. Hinton, C. Stegmann and F. A. Aharonian, \APJ{656}{2007}{870}
\bibitem{Beacom07}
J. F. Beacom and M. D. Kistler, \PRD{75}{2007}{083001}
\bibitem{Honda95}
M. Honda, T. Kajita, K. Kasahara and S. Midorikawa, \PRD{52}{1995}{4985}
\bibitem{Scanlon65}
J. H. Scanlon and S. N. Milford, \APJ{141}{1965}{718S}
\bibitem{Volkova88}
L. V. Volkova,  in "Erice 1988, Proceedings, Cosmic gamma rays, neutrinos, and related astrophysics". 
\bibitem{Barr88}
S. M. Barr, T. K. Gaisser, P. Lipari and S. Tilav, \PLB{214}{1988}{147}
\bibitem{Huang07ICRC}
C.-Y. Huang and M. Pohl, \ICRCYr{30}{2007}
\bibitem{Amato06}
E. Amato and P. Blasi, \MNRAS{371}{2006}{1251}
\bibitem{Berezhko99}
E. G. Berezhko and D. C. Ellison, \APJ{526}{1999}{385}
\bibitem{Wang97}
Z. R. Wang, Q.-Y. Qu and Y. Chen, \AA{318}{1997}{L59}
\bibitem{Lazendic04}
J. S. Lazendic, P. O. Slane, B. M. Gaensler et al., \APJ{602}{2004}{271}
\bibitem{Green01}
D.A. Green, \AIP{558}{2001}{59}
\bibitem{Crocker02}
R. M. Crocker, F. Melia and R. R. Volkas, \APJSuppl{141}{2002}{147} 
\bibitem{Aharonian07VelaJr}
The HESS Collaboration, F. A. Aharonian et al., \APJ{661}{2007}{236}
\bibitem{Kistler06}
M. D. Kistler and J. F. Beacom, \PRD{74}{2006}{063007}
\bibitem{Gandhi96}
R. Gandhi, C. Quigg, M. H. Reno and I. Sarcevic, \AstroPartPhys{5}{1996}{81}
\bibitem{Gandhi98}
R. Gandhi, C. Quigg, M. H. Reno and I. Sarcevic, \PRD{58}{1998}{093009}
\bibitem{Volkova80}
L. V. Volkova, \SovJNuclPhys{31}{1980}{784}
\bibitem{Ranft95}
J. Ranft, \PRD{51}{1995}{64}
\bibitem{Knapp03}
J. Knapp et al., \AstroPartPhys{19}{2003}{77}
\bibitem{Antoni01}
T. Antoni et al., \JPhysGNuclPartPhys{27}{2001}{1785}
\bibitem{Kasahara02}
K. Kasahara et al., \PRD{66}{2002}{052004}

\end{thebibliography}
\end{document}